\documentclass[12pt,preprint]{aastex}

\shortauthors{Spoon et al.}

\begin{document}

\title{Discovery of strongly blue shifted mid-infrared [Ne {\sc iii}]
and  [Ne {\sc v}] emission in ULIRGs}

\author{H.W.W. Spoon\altaffilmark{1}}
\email{spoon@isc.astro.cornell.edu}
\author{J. Holt\altaffilmark{2}}

\altaffiltext{1}{Cornell University, Astronomy Department, Ithaca, NY 14853}
\altaffiltext{2}{Leiden University, P.O. Box 9513, 2300 RA Leiden, The Netherlands}

\begin{abstract}
We report the discovery of blue shifted ($\Delta$v$>$200\,km s$^{-1}$)
mid-infrared [Ne{\sc iii}]  and/or [Ne{\sc v}] emission in 25 
out of 82 ULIRGs (30\% of our sample). The incidence of blue shifted 
[Ne{\sc v}] emission is even higher (59\%) among the sources 
with a [Ne{\sc v}] detection --- the tell-tale signature of 
an active galactic nucleus (AGN).
Sixteen ULIRGs in our sample, eleven of which are optically 
classified as AGN, have [Ne{\sc iii}] blue shifts above
200\,km s$^{-1}$. A comparison of the line profiles of their 
12.81\,$\mu$m [Ne{\sc ii}], 15.56\,$\mu$m [Ne{\sc iii}] and 
14.32\,$\mu$m [Ne{\sc v}] lines reveals the ionization of 
the blue shifted gas to increase with blue shift, implying 
decelerating outflows in a stratified medium, photo-ionized 
by the AGN.
The strong correlation of the line width of the [Ne{\sc iii}]
line with the radio luminosity indicates that interaction of 
expanding radio jets with the dense ISM surrounding the AGN
may explain the observed neon line kinematics for the strongest 
radio sources in this sample. 
\end{abstract}


\section{Introduction}

Galactic-scale outflows are now generally recognized to be 
important phenomena affecting or even regulating the evolution
of galaxies. Starburst-driven winds, powered by the combined 
kinetic energy released in the outflows from massive stars and 
supernovae, are capable of depositing metals, dust and energy 
into the galaxy halo and beyond \citep[][and references therein]{veilleux05}.
In galaxies hosting an active galactic nucleus (AGN), radiation-driven
AGN winds and collimated outflows of radio plasma, or ``jets'', 
offer another effective mechanism of transporting large amounts 
of energy and momentum out of the nucleus. In recent years, 
both processes have been implied in disrupting/fragmenting 
the obscuring cocoon in merger remnants \citep{hopkins05,holt08}.

In ultraluminous infrared galaxies (ULIRGs; 
L$_{\rm IR}$=10$^{12-13}$\,L$_{\odot}$), 
molecular gas driven into the merging nuclei by the galaxy
interaction is capable of supporting both a prolonged massive 
starburst and/or feeding a (nascent) AGN. This makes ULIRGs 
prime candidates for observing extreme AGN and/or starburst 
driven outflows. Indeed, observations of the interstellar
Na {\sc i}-D absorption line \citep[e.g.][]{heckman00,martin05,rupke05}
and optical forbidden low and medium ionization lines 
\citep[e.g.][]{heckman90,wilman99,holt03,lipari03} indicate
that outflows are ubiquitous in ULIRGs and that the momentum
and energy injection is in most cases dominated by the starburst
\citep{rupke05,veilleux05}.

Mid-infrared kinematic studies of forbidden line emission in AGN and 
ULIRGs are scarce. In their ISO-SWS survey of local AGN, \cite{sturm02}
reported the detection of line asymmetries and line shifts for some
of their sources. Detailed studies could, however, only be performed 
for the two brightest AGNs, NGC\,4151 \citep{sturm99} and NGC\,1068 
\citep{lutz00}. For these it was determined that dust entrained
in outflows cannot explain the observed asymmetries in optical 
forbidden lines. 
With the advent of Spitzer-IRS sensitive medium-resolution 
spectroscopy has become available for kinematic studies of AGN 
and ULIRGs at a resolving power $R$=600. \cite{dasyra08} have exploited 
these capabilities and found the line widths of the high-ionization
AGN narrow line tracers 14.32\,$\mu$m [Ne{\sc v}] and 25.89\,$\mu$m 
[O{\sc iv}] to be correlated with the black hole mass for AGNs
for which the narrow line region is dominated by virial motion.
More recently, \cite{spoon09} presented the discovery of strongly
blueshifted ionized Ne$^+$, Ne$^{2+}$ and Ne$^{4+}$ gas in three 
ULIRGs and proposed that the outflowing gas may be tracing the 
disruption of the obscuring medium around buried AGNs in these mergers.  
In this Letter we report the mid-infrared discovery of strongly 
blue shifted neon line emission in another 22 ULIRGs. 



\section{Observations and data reduction}

The results presented in this paper are based on high resolution 
($R$$\sim$600; $\Delta$$v\sim$500\,km s$^{-1}$) 10--37\,$\mu$m Spitzer 
\citep{werner04} IRS \citep{houck04} spectra of ULIRGs, AGN and 
starburst galaxies. The core sample is formed by the IRS GTO sample 
\citep{weedman05,farrah07,bernard-salas09} which is not complete 
in any sense.

All sources were observed in staring mode. The 10--19.5\,$\mu$m
portion of the spectra were obtained in the IRS Short High (SH)
module and the 19.3--37\,$\mu$m sections in the Long High (LH) 
module. The data reduction proceeded along 
the same steps as detailed in \cite{spoon09}.
The wavelength calibration of our spectra should be 
accurate to about 1/5 (SSC priv. comm.) of a resolution element 
(500 km s$^{-1}$), which amounts to 100 km s$^{-1}$. The accuracy
is however known to vary from order to order and is generally
lowest near the order edges and best toward the order center.

\section{Analysis}

We have used the non-linear least squares fitting IDL package
MPFIT \citep{markwardt08} to perform single and double component 
gauss fits to the high-resolution Spitzer emission line spectra 
of the galaxies in our sample. With these fits we computed the 
full width and velocity center at 20\% of the peak flux (FW20
and VC20, respectively) for all detected lines. Our choice of 
evaluating at the 20\% level rather than at the 50\% level 
is motivated by the presence of weak blue wings in the neon 
lines.

In this particular study we focus on the mid-infrared fine 
structure lines of neon, because of their large range in 
ionization potential (IP = 21, 41, 97 and 127\,eV for Ne$^+$, 
Ne$^{2+}$, Ne$^{4+}$ and Ne$^{5+}$, respectively) and their 
superior detectability compared to fine-structure lines of 
argon and sulphur. Also, the rest wavelengths of the five 
neon lines are well away from the strong Si--O stretching 
resonance of amorphous silicate grains at 8--12$\mu$m which 
is responsible for a factor 3--5 increase in opacity compared 
to surrounding wavelengths. Furthermore, the three strongest 
lines, 12.81\,$\mu$m [Ne {\sc ii}], 15.56\,$\mu$m [Ne {\sc iii}] 
and 14.32\,$\mu$m [Ne {\sc v}], all fall within the same IRS slit, 
SH, for redshifts up to 0.25. At higher redshifts the 15.56\,$\mu$m 
[Ne {\sc iii}] line shifts from the SH slit into the LH slit, 
while above z=0.32 the 7.65\,$\mu$m [Ne {\sc vi}] line shifts 
into the SH slit. 
Our neon line sample consists of 82 ULIRGs, 3 HyLIRGs,
33 AGNs (both Seyferts and QSOs) and 15 starburst galaxies 
suited for detailed line profile analysis. 

\begin{figure}
\begin{center}
\includegraphics[angle=0,width=8.8cm]{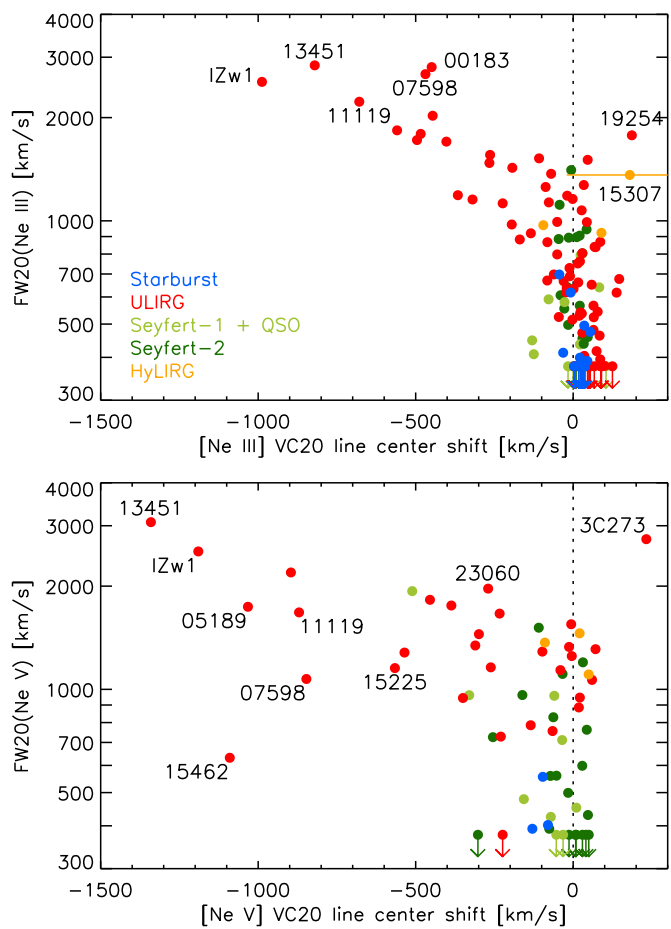}
\caption{The full width at 20\% of the peak flux (FW20) for
the line profiles of 15.56\,$\mu$m [Ne {\sc iii}] and 
14.32\,$\mu$m [Ne {\sc v}] lines shown as a function of the 
shift of the line center measured at 20\% of the peak height 
(VC20). Typical uncertainties for the FW20 values are 10\% 
and 15\% for the [Ne {\sc iii}] and [Ne {\sc v}] lines, 
respectively. Upper limits on FW20 values denote unresolved lines.
The sources are color-coded as follows:
ULIRGs are {\it red}, starburst galaxies {\it blue}, QSOs and
Seyfert-1 galaxies {\it light green}, and Seyfert-2 galaxies 
{\it dark green}. 
\label{fig1}}
\end{center}
\end{figure}

In  Fig.\,\ref{fig1} we plot for the [Ne {\sc iii}] 
and [Ne {\sc v}] lines the FW20 as a function of VC20 for 
all sources in our sample. The panels strikingly illustrate 
the predominance of blue over red shifts among the neon line
detections, and the tendency for the VC20 blue shift to
be correlated with the FW20 line width. The latter appears
to be best established for the [Ne {\sc iii}] line. 

Among our ULIRG sample 25/82 sources (30\%) show blue shifted
[Ne {\sc iii}] and/or [Ne {\sc v}] emission. This fraction 
excludes sources with VC20 blue shifts less than 200\,km s$^{-1}$, 
for which we cannot rule out uncertainties in the order-to-order 
wavelength calibration to mimick a real blue shift. 21/25 ULIRGs 
(84\%) are optically classified as AGN.
Among the 32/82 ULIRGs with a [Ne {\sc v}] detection, 
the [Ne {\sc v}] line is found to be blue shifted 
(VC20$<$-200\,km s$^{-1}$) for 19 ULIRGs (59\%), 
6 of which are optically classified 
as type-{\sc i} and 10 as type-{\sc ii} AGN.

In Fig.\,\ref{fig1} the broadest [Ne{\sc v}] lines 
not showing clear indications for blue shifts have FW20 
line widths of 1500\,km s$^{-1}$. Assuming these lines
to arise in NLR gas in keplerian rotation around
the central black hole, equation 1 of \cite{dasyra08}
predicts their black hole masses to be of the order of
10$^{9.3}$\,M$_{\odot}$.

In Figs.\,\ref{fig2} and \ref{fig3} we show overlays of the 
velocity profiles of the mid-infrared neon lines for 16 sources 
which have [Ne{\sc iii}] VC20 line center blue shifts of 
200\,km s$^{-1}$ or more. The sources are sorted by decreasing 
FW20 of the [Ne{\sc iii}] line. Eleven of these are optically
classified as AGN. 
The 16 panels reveal a great diversity in line shapes and line 
center shifts, both from line to line for a single target, as 
well as from object to object. In some sources the VC20 line center
shift is the result of a pronounced blue wing to the systemic 
component, while in others also the line peak is blue shifted. 
The latter is more often the case for the [Ne {\sc v}] line
than for the [Ne {\sc iii}] line, and can also be more extreme. 
For other sources two gaussian components can be discerned,
one at systemic velocity, the other at blue shifted velocities. 
Examples are the [Ne {\sc ii}] line of IRAS\,07598+6508 and 
the [Ne {\sc iii}] line of IRAS\,15462--0450. 
In Table\,\ref{tab1} we list the measurements of the kinematic 
parameters for the sources in Figs.\,\ref{fig2} and \ref{fig3}
along with their computed 8--1000\,$\mu$m infrared luminosities 
and 1.4\,GHz radio luminosities.

With lines of various ionization stages of neon measured
in the same aperture, it is possible to investigate
how the ionization of the gas changes as a function of velocity
relative to systemic. In Figs.\,\ref{fig2} and \ref{fig3} we 
show the change in [Ne {\sc iii}]/[Ne {\sc ii}] and 
[Ne {\sc v}]/[Ne {\sc iii}]
line ratios at velocity intervals of 750\,km s$^{-1}$ with 
respect to systemic. Results are only 
shown for those velocity bins for which the ratio-ed lines 
are at least 1/20 of the peak line flux throughout the 
central 600\,km s$^{-1}$ of the velocity bin. 
For nearly all the sources 
for which either line ratio is defined over more than one 
velocity bin, the ionization appears to increase with 
increasing blue shift. In some cases this can be substantial, 
like for IZw1, where the [Ne {\sc v}]/[Ne {\sc iii}] ratio 
increases from 0.21 at systemic velocity to 1.3 in the 
velocity bin centered at -1500\,km s$^{-1}$. A clear exception
is IRAS\,F00183--7111, for which the [Ne {\sc iii}]/[Ne {\sc ii}]
ratio is remarkably constant over a wide velocity range
\citep{spoon09}.
The above trends can also be recognized in a more visual way by 
comparing for each source which neon lines have the most pronounced 
blue wings. This reveals that for the large majority of sources 
in Figs.\,\ref{fig2} and \ref{fig3} the blue wing of the [Ne {\sc v}] 
line extends beyond that of the [Ne {\sc iii}] line, while the 
blue wing of the [Ne {\sc iii}] line extends beyond that of 
the [Ne {\sc ii}] line; consistent with an increase in ionization 
with increasing blue shift.

All sources in Figs.\,\ref{fig2} and \ref{fig3} display an 
unresolved or barely resolved [Ne {\sc ii}] component at 
systemic velocity. Some of these sources also have a [Ne {\sc iii}]
or even [Ne {\sc v}] systemic component. The excitation of
this neon gas holds clues as to the nature of the dominant 
nuclear ionizing source.
In pure starburst galaxies the excitation of the nuclear
starburst as measured from the [Ne {\sc iii}]/[Ne {\sc ii}] 
line ratio is usually low and in the range of 0.05--0.2 
\citep{thornley00,bernard-salas09}\footnote{A notable exception 
may be the pure starburst galaxy NGC\,7714, which has a 
measured [Ne {\sc iii}]/[Ne {\sc ii}] ratio of 0.75 
\citep{bernard-salas09}}. In our sample of sixteen excitation 
conditions like these are met for the systemic components
of IRAS\,05024--1941, IRAS\,07598+6508, IRAS\,12127--1412NE, 
IRAS\,15462--0450 and IRAS\,15225+2350. Among these 
IRAS\,15225+2350 is the only source optically classified 
as a starburst.
For other sources, like IRAS\,12514+1027 and IRAS\,16156+0146NW,
the [Ne {\sc iii}]/[Ne {\sc ii}] ratios for the velocity component
at systemic velocity are far higher, 0.97 and 0.78, respectively.
In these galaxies the AGN contributes to the systemic
[Ne {\sc iii}] emission, given the detection of a systemic
[Ne {\sc v}] emission component.
For IRAS\,05189--2524 and IRAS\,11119+3257 the clear blue shift 
of both the [Ne{\sc iii}] and [Ne{\sc v}] emission implies that 
none of the higher ionization gas in these sources has a keplerian 
component.

In Fig.\,\ref{fig4} we plot the FW20 line width of the [Ne {\sc iii}]
and [Ne {\sc v}] lines as a function of 1.4\,GHz radio luminosity for 
all AGNs and ULIRGs in our sample. This reveals a clear trend of 
increasing line width with increasing radio power. A similar
trend was first found for the 5007\,\AA\ [O {\sc iii}] line
by \cite{wilson80} and \cite{whittle85} for samples of radio 
quiet (L[1.4GHz]$<$10$^{23}$\,W Hz$^{-1}$) seyfert galaxies and 
QSOs. In our data the trend appears to persist -- albeit with 
some outliers -- well into the radio-loud domain. The sources
with strong blue shifts in their [Ne{\sc iii}] lines, all labeled 
in red, appear to define the upper envelope of this trend or
a separate one. This is especially clear in the upper panel. 
The only obvious exception is the prototypical narrow-line 
Seyfert 1 (NLS1) galaxy IZw1. 
Note that IRAS\,07598+6508 is not included, as it lacks a 1.4\,GHz 
flux measurement.


\section{Discussion}

\begin{figure}
\begin{center}
\includegraphics[angle=0,width=8.8cm]{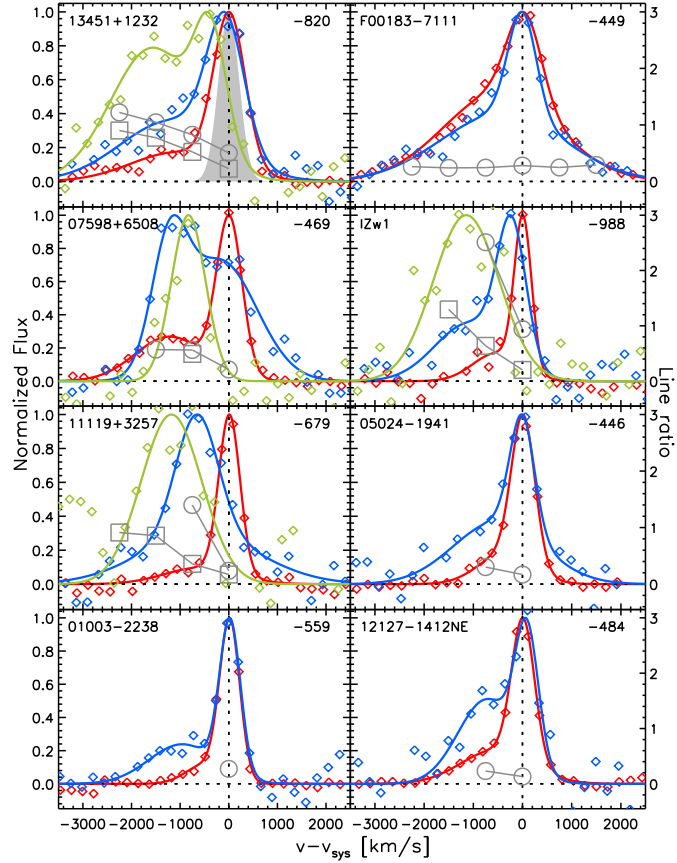}
\caption{Comparison of the line profiles of 12.81\,$\mu$m [Ne {\sc ii}]
({\it red}), 15.56\,$\mu$m [Ne {\sc iii}] ({\it blue}) and 14.32\,$\mu$m
[Ne {\sc v}] ({\it green}), as observed with the Spitzer-IRS 
high-resolution spectrographs at $R$=600. The sources are sorted 
in order of decreasing FW20 of the [Ne{\sc iii}] line.
Systemic velocity is indicated by vertical {\it dashed} lines.
Data points are shown as {\it diamonds}, fits to the data points 
are shown as {\it continuous} lines.
The {\it filled gray area} in the top-left panel shows the gauss 
profile of an unresolved emission line at a resolution of $R$=600.
Also plotted are the line ratios of [Ne {\sc iii}]/[Ne {\sc ii}]
({\it gray open circles}) and [Ne {\sc v}]/[Ne {\sc iii}]
({\it gray open squares}) at 750\,km s$^{-1}$ velocity intervals.
These ratios are computed from the fitted line profiles. The number
in the upper right corner is the VC20 line shift of the [Ne {\sc iii}]
line in km s$^{-1}$.
\label{fig2}}
\end{center}
\end{figure}

\begin{figure}
\begin{center}
\includegraphics[angle=0,width=8.8cm]{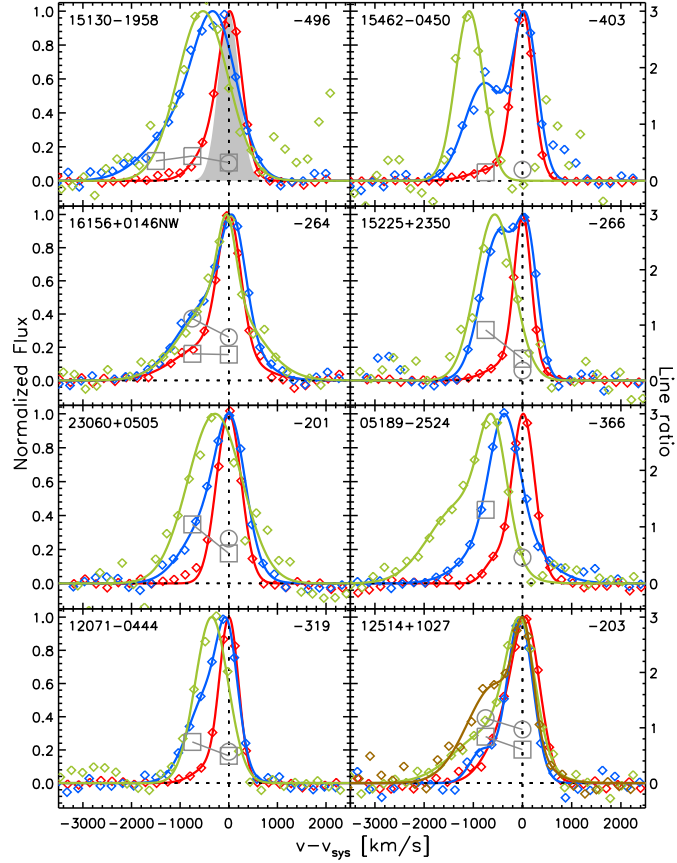}
\caption{Comparison of the line profiles of 12.81\,$\mu$m [Ne{\sc ii}]
({\it red}), 15.56\,$\mu$m [Ne{\sc iii}] ({\it blue}), 14.32\,$\mu$m
[Ne{\sc v}] ({\it green}) and 7.65\,$\mu$m [Ne{\sc vi}] ({\it brown}; 
only IRAS\,12514+1027), as observed with the Spitzer-IRS high-resolution 
spectrographs at $R$=600. The lay-out is the same as in Fig.\,\ref{fig2}.
\label{fig3}}
\end{center}
\end{figure}

A striking result of our study is the 
identification of a sizeable subsample of sources which 
show a strong blue shift of the [Ne{\sc v}] line center 
in combination with a smaller blue shift of the [Ne{\sc iii}] 
line center (Figs.\,\ref{fig2} and \ref{fig3}) in their spectra.
A per-source comparison of the [Ne{\sc ii}], [Ne{\sc iii}] 
and [Ne{\sc v}] profiles strongly suggests these shifts to 
be the result of an increase in ionization with increasing 
blue shift in the line. This is supported by the behavior 
of the [Ne{\sc iii}]/[Ne{\sc ii}] and  [Ne{\sc v}]/[Ne{\sc iii}] 
line ratios as a function of line velocity. The most extreme
sources in this subsample do not even exhibit a [Ne{\sc v}] 
component at systemic velocity. Examples are IZw1, 
IRAS\,05189--2524, IRAS\,07598+6508, IRAS\,11119+3257 and 
IRAS\,15462--0450. This strongly suggests that at least 
in these sources the observed [Ne{\sc v}] emission 
cannot arise in NLR gas in keplerian rotation. Instead, 
the observed line shifts and blue wings imply a nuclear
outflow in a highly stratified ISM, ionized by the central 
source \citep[e.g. IRAS\,13451+1232;][]{spoon09}. The observed 
increase of ionization with increasing blue shift then 
indicates that the bulk of the observed neon gas is moving 
toward us, and that the outflow speed decreases with distance 
to the ionizing source.
The absence of a matching {\it red}ward line component
to the [Ne{\sc v}] line implies that the line of sight 
to the far side of the outflow is blocked at mid-infrared 
wavelengths.

Note that most of these sources are also known from optical 
spectroscopy to exhibit outflows. Nine of the 16 sources 
in Table\,\ref{tab1} are listed by \cite{lipari03} as 
showing optical outflow signatures, but also IZw1 and
IRAS\,23060+0505 have known optical outflows 
\citep{laor97,wilman99}.

Results similar to ours were recently reported for a 
sample of ``blue outlier'' NLS1 galaxies \citep{komossa08}.
For these galaxies a similar correlation seems to 
exist between blue shift and line width
of the 5007\,\AA\ [O{\sc iii}] line as we see for the
15.56\,$\mu$m [Ne{\sc iii}] line (and for the 
14.32\,$\mu$m [Ne{\sc v}] line to a lesser degree).
The authors favor a scenario in which gas clouds from
the narrow line region are entrained in a decelerating
wind, powered either by collimated radio plasma jets,
or radiation pressure resulting from accretion close
to or above the Eddington limit. 
Similar [O{\sc iii}] blue shifts and line widths have 
also been reported for two young dust-enshrouded radio 
galaxies: PKS\,1549--79 \citep{tadhunter01} and 
IRAS\,13451+1232 \citep{holt03}. Both are late stage
mergers and the latter is one is part of our sample
(Table\,\ref{tab1}). The broad, blue shifted [O{\sc iii}] 
emission from these sources may trace the expansion
of radio jets through the obscuring shell around the
buried AGN \citep{holt08}.

\begin{figure}
\begin{center}
\includegraphics[angle=0,width=8.8cm]{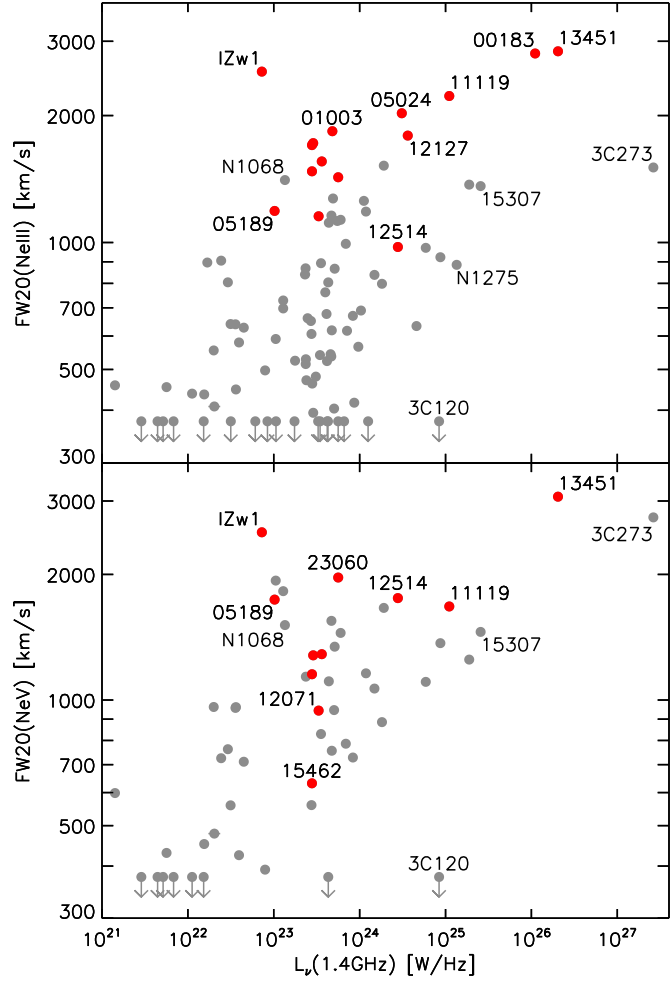}
\caption{The full width at 20\% of the peak flux (FW20) 
for the 15.56\,$\mu$m [Ne {\sc iii}] and the
14.32\,$\mu$m [Ne {\sc v}] line profile shown as a 
function of the 1.4\,GHz radio luminosity density. Typical 
uncertainties for the FW20 values are 10\% and 15\%, 
respectively. 
Upper limits on FW20 values denote unresolved lines.
The sources shown in red all have  [Ne {\sc iii}] VC20
blue shifts of 200\,km s$^{-1}$ or higher and are the
same sources as listed in Table\,\ref{tab1}
\label{fig4}}
\end{center}
\end{figure}

Given the high radio luminosities for IRAS\,13451+1232,
IRAS\,F00183--7111, IRAS\,11119+3257, IRAS\,12127--1412NE,
and IRAS\,05024--1941, the apparent correlation between radio 
luminosity and [Ne{\sc iii}] line width in Fig.\,\ref{fig4} 
may indicate that for these sources interaction of 
expanding radio jets with the dense ISM surrounding 
the AGN may explain the observed neon line kinematics.

Four of the sources in Fig.\,\ref{fig2} which show outflow
signatures in their neon lines do not show the tell-tale 
mid-infrared signature of AGN activity, the 
presence of a 14.32\,$\mu$m  [Ne{\sc v}] line, in their 
spectrum. While this may indicate their outflows to be
starburst-driven, three of these sources, IRAS\,F00183--7111,
IRAS\,05024--1941 and IRAS\,12127--1412NE, display strong 
evidence for the presence of a powerful AGN -- e.g. 1.4\,GHz 
radio fluxes 4--10 times in excess of the value expected 
for star forming galaxies. 
In addition, IRAS\,F00183--7111 and IRAS\,12127--1412NE 
have silicate strengths \citep[S$_{sil}$;][]{spoon07} 
below -2.5, indicating an apparent silicate optical depth of 
at least 2.5 toward the nucleus. Hence, it is conceivable that
the non-detection of [Ne{\sc v}] emission in these sources 
is the result of obscuration (along our line of sight)
of the interior, most highly ionized part of the outflow 
rather than the absence or insignificance of an AGN in 
these sources.
For the fourth galaxy, IRAS\,01003--2238, conflicting 
classifications exist. Several studies have concluded 
that this galaxy may be starburst-powered 
\citep{armus88,rigopoulou99,lutz99,wilman99}, whereas
\cite{allen91} classify this galaxy as Seyfert-II
and \cite{lipari03} refer to it as a QSO. Its Spitzer-IRS
spectrum is continuum-dominated \citep[class 1A;][]{spoon07}
with a 6.2\,$\mu$m PAH equivalent width more than ten
times lower than for starburst galaxies. From the PAH
luminosity we calculate a contribution of the starburst
to the infrared luminosity of 10--20\% \citep{peeters04}.
This would imply that IRAS\,01003--2238 is AGN-dominated\footnote{The non-detection 
of the 14.32\,$\mu$m [Ne{\sc v}] line and of the 
near-infrared AGN tracer 1.962\,$\mu$m [Si{\sc vi}]
\citep{veilleux97,dannerbauer05} is not exceptional 
among AGN-dominated ULIRGs. Other examples are
Mrk\,231 \citep{armus07} and IRAS\,00275--2859} and 
its outflow likely AGN-powered.
The mere detection of blue wings in [Ne{\sc ii}] and 
[Ne{\sc iii}] line profiles in Spitzer-IRS-SH galaxy 
spectra may hence be a general signpost of AGN activity.


\acknowledgements

The author thanks Ian Waters for help with measuring the line profiles
and Jeremy Darling, Brent Groves, James Houck and Stefanie Komossa for 
discussions and Riccardo Giovanelli for the well-timed suggestion to 
publish this result as a Letter.
This work is based on observations made with the {\it Spitzer Space Telescope},
which is operated by the Jet Propulsion Laboratory (JPL), California
Institute of Technology, under NASA contract 1407. Support for this
work by the IRS GTO team at Cornell University was provided by NASA
through contract 1257184 issued by JPL, Caltech.


\clearpage

\begin{deluxetable}{lrlrrrrrrrr}
\tabletypesize{\scriptsize}
\rotate
\tablecolumns{11}
\tablewidth{0pc} 
\tablecaption{Neon emission line kinematics\label{tab1}}
\tablehead{\colhead{Target name} &  \colhead{z\tablenotemark{a}} & 
           \colhead{class\tablenotemark{b}} &
           \colhead{FW20$_{\rm [NeII]}$} & \colhead{FW20$_{\rm [NeIII]}$} &  
           \colhead{FW20$_{\rm [NeV]}$} & \colhead{VC20$_{\rm [NeII]}$} &
           \colhead{VC20$_{\rm [NeIII]}$} & \colhead{VC20$_{\rm [NeV]}$} &
           \colhead{log L(IR)\tablenotemark{c}} &
           \colhead{log L(1.4\,GHz)\tablenotemark{d}}\\
           \colhead{(IRAS)} &  \colhead{} & \colhead{} &
           \colhead{(km s$^{-1}$)} & \colhead{(km s$^{-1}$)} &  
           \colhead{(km s$^{-1}$)} & \colhead{(km s$^{-1}$)} &
           \colhead{(km s$^{-1}$)} & \colhead{(km s$^{-1}$)} &
           \colhead{(L$_{\odot}$)} & \colhead{(W Hz$^{-1}$)}
}
\startdata
00183--7111  &0.3219 &L/A (1,2) &  3037&2807&   --& -393& -449&   --&12.93&26.04 (1)\\
IZw1         &0.06089&NLS1 (3)&$<$400&2543& 2524&    0& -988&-1190&11.95&22.86 (2)\\
01003--2238  &0.1179 &H/S2 (4,7) &$<$400&1837&   --&    0& -559& --&12.29&23.68 (2)\\
05024--1941  &0.1937 &S2 (5)&   773&2026&   --&  -66& -446&   --&12.39&24.49 (2)\\
05189--2524  &0.04274&S2 (5)&   580&1189& 1741&    0& -366&-1032&12.16&23.01 (3)\\
07598+6508   &0.1484 &S1\tablenotemark{e} (5)&  2109&2678& 1073& -626& -469&-847&12.53&--\\
11119+3257   &0.1903 &S1 (5)&   473&2226& 1677&    0& -679& -870&12.66& 25.04 (2)\\ 
12071--0444  &0.1286 &S2 (5)&$<$400&1155&  945&  -23& -320& -350&12.39& 23.52 (4)\\
12127--1412NE&0.1330 &L (5) &   986&1794&   --& -146& -484&   --&12.17& 24.56 (2)\\
12514+1027   &0.3194 &S2 (6)&   969& 977& 1756&    0& -203& -387&12.77& 24.44 (2)\\
13451+1232   &0.1214 &S2 (5)&  1228&2841& 3072& -109& -821&-1340&12.29& 26.31 (2)\\
15130--1958  &0.1103 &S2 (5)&   702&1721& 1281&    0& -496& -535&12.14& 23.46 (2)\\
15225+2350   &0.1386 &H (5) &$<$400&1476& 1154&  -14& -266& -566&12.16& 23.44 (4)\\
15462--0450  &0.1001 &NLS1 (3)&$<$400&1703&  634&  -16& -403&-1090&12.20& 23.44 (4)\\  
16156+0146NW &0.1326 &S2 (5)&   943&1558& 1289& -109& -264&  -98&12.08& 23.56 (4)\\
23060+0505   &0.1734 &S2/L (6)&   524&1429& 1966&    0& -201&-270&12.53& 23.75 (4)\\
\enddata
\tablecomments{Results from the 1 or 2-component gauss
fits to the 12.81 [Ne{\sc ii}], 15.56 [Ne{\sc iii}] and 14.32 [Ne{\sc v}] 
line profiles. FW20 is defined as the full width at 20\% of the peak flux, 
and is corrected for the instrumental profile.
We estimate the uncertainty in these quantities to be 5, 10 and 15\%
for the [Ne{\sc ii}], [Ne{\sc iii}] and [Ne{\sc v}] line, respectively.
VC20 is the velocity shift of the line center measured at 20\% of the
peak flux. The uncertainty in the VC20 values of the neon lines are 
limited to order-to-order wavelength calibration uncertainties which 
we estimate to be smaller than 150\,km s$^{-1}$.
}
\tablenotetext{a}{The adopted redshift ($z$) is defined by the 
line center of the [Ne{\sc ii}] line measured at 80\% of the peak flux (VC80).
}
\tablenotetext{b}{Optical spectral classifications: (S1) Seyfert type 1, 
(NLS1) narrow-line Seyfert type 1,
(S2) Seyfert type 2, (L) LINER, (H) H{\sc ii}, (A) Ambiguous.
Classifications were taken from (1) \cite{armus89},
(2) \cite{buchanan06}, (3) \cite{veron-cetty06}, (4) \cite{veilleux99}, 
(5) \cite{veilleux02}, (6) \cite{wilman99}, (7) \cite{allen91} 
}
\tablenotetext{c}{L(IR)=L(8--1000\,$\mu$m) as defined by \cite{sanders96}}
\tablenotetext{d}{Radio luminosities were compiled from: (1) 
\cite{drake04}, (2) NVSS \citep{condon98}, (3) \cite{yun01}, (4) FIRST \citep{white97}
}
\tablenotetext{e}{Broad absorption line (BAL) QSO \citep{lipari94}
}
\end{deluxetable}

\clearpage

\clearpage

\clearpage

\clearpage

\end{document}